# The architecture of co-culture spheroids regulates tumor invasion within a 3D extracellular matrix

Yu Ling Huang[a], Carina Shiau [a], Cindy Wu [a], Jeffrey E. Segall [b], and Mingming Wu [a*]

[a] Department of Biological and Environmental Engineering, 306 Riley-Robb Hall, Cornell University, Ithaca, NY 14853

[b] Anatomy and Structural Biology, Albert Einstein College of Medicine, 1300 Morris Park Avenue, Bronx, New York 10461

*Corresponding author: mw272@cornell.edu

## Abstract

Tumor invasion, the process by which tumor cells break away from their primary tumor and gain access to vascular systems, is an important step in cancer metastasis. Most current 3D tumor invasion assays consisted of single tumor cells embedded within an extracellular matrix (ECM). These assays taught us much of what we know today on how key biophysical (e.g. ECM stiffness) and biochemical (e.g. cytokine gradients) parameters within the tumor microenvironment guided and regulated tumor invasion. One limitation of the single tumor cell invasion assay was that it did not account for cell-cell adhesion within the tumor. In this article, we developed a micrometer scale 3D co-culture spheroid invasion assay that was compatible with microscopic imaging. Micrometer scale co-culture spheroids (1:1 ratio of metastatic breast cancer MDA-MB-231 and non-tumorigenic epithelial MCF-10A cells) were made using an array of microwells, and then were embedded within a collagen matrix in a microfluidic platform. Real time imaging of tumor spheroid invasion revealed that the spatial distribution of the two cell types within the tumor spheroid critically regulated tumor invasion. This work linked tumor architecture with tumor invasion and highlighted the importance of the biophysical cues within the bulk of the tumor in tumor invasion.

**Introduction**

Visual inspection of cell/nucleus shape along with spatial distribution of different cell types has always been a critical component of cancer diagnosis. An important method of identifying tumor progression stage is to inspect a 2D slice of tumor tissue under a microscope, with infiltration of immune cells and large nucleus size indicating a poor prognosis [1]. For breast cancer, the cell-cell spatial arrangement, or the architecture of the tumor, has been used for diagnosing tumor invasiveness. An early form of breast cancer is ductal carcinoma in situ (DCIS) where the presence of abnormal cells are inside the milk duct [2, 3]. Viewing from a cross-section, DCIS can be described as a hollow circular epithelial cell shell with abnormal cells at the center of the shell. DCIS is typically noninvasive, however, it can be transformed into an invasive form when the abnormal cells break away from the normal epithelial cell layer and invade into the surrounding tissue [4, 5].

Tumor invasion is an important step of cancer metastasis because tumor cells need to break away from the primary tumor to invade into their surrounding environment [6, 7]. Extensive work has been carried out demonstrating that cell/nucleus shape and mechanics have been correlated with the invasiveness of the tumor cells using assays involving single tumor cells plated on a 2D substrate or single tumor cells embedded within a 3D extracellular matrix (ECM)[8-12]. In vivo, solid tumors often are in the form of cell aggregates or compact cell mass containing many cell types including normal epithelial, endothelial, immune and tumor cells [13]. The architecture of breast tumor is thought to evolve with time, from an early stage DCIS with a simple core shell structure to a late stage tumor with a complex architecture involving vasculature. There is an extensive literature in understanding the roles of the tumor microenvironment in promoting tumor invasion [6, 14-18]. However, less is known about how tissue architecture within the bulk of the tumor influences tumor invasion [19, 20].

Accordingly, we developed a co-culture tumor spheroid assay for studies of tumor architecture in tumor invasion that is compatible with optical microscopic imaging. We used 1:1 ratio of metastatic and non-tumorigenic epithelial cells to create co-culture tumor spheroids and a type I collagen matrix as the surrounding 3D ECM. The invasion behavior of the tumor spheroids was followed by time lapse microscopy. Our work demonstrated that tumor architecture critically regulated tumor cell invasion.

## Results and Discussion:

*Formation of co-culture tumor spheroids*

Co-culture spheroids were formed within an array microwell platform previously developed in our labs [21, 22]. 1:1 cell number ratio of metastatic MDA-MB-231expressing EGFP (green) and non-tumorigenic MCF-10A cells expressing dTomato (red) fluorescent protein were placed in the microwells. Upon seeding, the cells of two types mixed uniformly as seen in Fig. 1A. With time, cells of both types moved, proliferated, and re-organized within the microwell, and tumor spheroids started to form after overnight incubation. Interestingly, we found that the spatial organization of the two cell types within the co-culture spheroids evolved over time. On day 2, the non-tumorigenic cells (red MCF-10A cells) formed a cluster in the center with metastatic cancer cells (green MDA-MB-231 cells) on the periphery (top panels of Fig. 1B and C). On day 4, this spatial organization was reversed with MCF-10A cells outside and MDA-MB-231 cells inside (low panels of Fig. 1B and C). The degree of enclosure of one cell type over the other was not uniform from well to well, and varied as indicated in Figure 1C.

The phenomena of cell segregation has been studied extensively using differential cell-cell adhesion theory [23, 24], however, spheroid architecture inversion was discovered only recently and was poorly understood . We know that the malignant MDA-MB-231 cells are transformed cells, and have essentially no or very low expression of the cell-cell adhesion molecule E-cadherin. MCF-10A cells, on the other hand, have high E-cadherin expression [21, 24]. During spheroid formation, cells of two different types can migrate and re-organize to keep the total free energy at the lowest level. Here the free energy is computed using cell-cell adhesion. In addition, cells of both types proliferate and have different growth rates. In these conditions, MCF-10A cells have a higher proliferation rate than MDA-MB-231 cells. Our previous studies showed that the differential growth rate was an important factor for the spheroid architecture inversion. No architecture inversion was observed when cell growth was inhibited [21]. In this work, we use this inversion phenomena to generate spheroids of different architecture, further experiments will be needed to understand the underlying mechanism responsible for the inversion.

*The architecture of co-culture spheroid regulated tumor invasion*

To examine how the architecture of the co-culture spheroids affects tumor cell invasion, we embedded the spheroids within a 3D matrix at a concentration of 1.5 mg/mL type I collagen. We placed the spheroid embedded ECM into a microfluidic platform developed previously in the lab [25, 26] and followed the spatial and temporal dynamics of the spheroids over a time course of 36 hours (Fig. 2). Here, we found distinct tumor invasion patterns using co-culture spheroids of four different architecture. In architecture S2a, where MDA-MB-231 cells surrounded the MCF-10A cells, a significant population of MDA-MB-231 cells were observed to detach from the spheroid (Fig. 2A and Movie S1). In contrast, in the spheroids with a reversed architecture (S4a), only a few MDA-MB-231 cells at the periphery were able to invade out, but majority of the MDA-MB-231 cells remained in the spheroid core enclosed by a MCF-10A shell (Fig. 2B and Movie S2). In architectures S2b and S4b, where part of the MDA-MB-231 cells was at the periphery and in direct contact with collagen matrix initially, majority of the tumor cells invaded out from the side of the spheroid away from the MCF-10A aggregate ((Fig. 2C and 2D, Movies S3 and S4). In particular, in the spheroids with architecture S4b where MDA-MB-231 cells were in contact with collagen through a small opening of the MCF-10A shell, MDA-MB-231 cells streamed out through the opening in a directional way (Fig. 2D). Fig. 2 clearly demonstrated that the initial cell-cell spatial arrangements within the co-culture spheroid significantly influenced the subsequent invasion.

*Malignant tumor cells detached from the spheroid more readily in day 2 spheroids than day 4 spheroids*

The first step of tumor cell invasion is for the malignant tumor cells to detach from the tumor spheroids. Using the time sequence images of tumor invasion, we quantified the number of malignant cells detached from the spheroid by using the fluorescence of the cells in the image. We assumed that the total fluorescence intensity of the tumor cells was proportional to the number of cells in this calculation. We found that MDA-MB-231 cells were more readily detached from the spheroids harvested at day 2 than those from spheroids harvested at day 4 (Fig. 3) regardless of their initial architecture. At t = 36 hours, our result demonstrated that spheroids S2a and S2b had the high percentages of detaching cells of 55.0 ± 1.19 % and 50.5 ± 0.88 %. In contrast, the percentages of detached cells from spheroids S4a and S4b were 22.1 ± 5.07 % and 34.9 ± 7.34 %. This is consistent with the understanding that spheroids harvested at Day 2 were not as compact as those from Day 4. We know that cells secret matrices and adhesion molecules during spheroid formation which may be the reason for spheroid compaction with time. In addition, the non-tumorigenic cells at the periphery secured the metastatic tumor cell cluster within the spheroid core and hence blocked their invasion.

*Invasion characteristics of malignant tumor cells from co-culture spheroids*

An important characteristic for tumor invasion is how far tumor cells migrate away from the primary tumor within a given time. To quantify the motility of tumor cells from co-culture spheroids of different architectures, we tracked individual MDA-MB-231 malignant breast tumor cells that invaded out from the spheroid. The trajectories of 60 cells for each of the four different spheroid architecture are shown in Figure 4. Here, the starting point of the track was the time when the cell detached from tumor spheroid, and end point of the track was marked with a dot.

Using the cell trajectories, we computed the cell migration speed, persistence, and mean squared displacements (MSDs) of the MDA-MB-231 cells under these four conditions. We first compared the cell motilities between S2a (MCF-10A cell cluster surrounded by MDA-MB-231 cells) and S4a (the reversed architecture of S2a, with loosely associated MDA-MB-231 cells around the MCF-10A spheroid shell) architectures. Surprisingly, there was no significant difference between the cell migration speed. The cell migration speed for S2a is $0.166 \pm 0.006$ μm/min, while the average cell migration speed for S4a is $0.164 \pm 0.007$ μm/min. However, the persistence of the tumor cells in S4a is greater than those in S2a, with an average persistence of $0.574 \pm 0.022$ for S4a in contrast to $0.491 \pm 0.021$ for S2a, or a 16.9% increase. Accordingly, there is a slight increase at later time points in the MSDs measured for S4a spheroids than S2a spheroids. We then compared the cell motilities between S4b (MDA-MB-231 cells with an exit route from full enclosure by MCF-10A cells) and S2b (MDA-MB-231 and MCF-10A cluster side by side) architectures. Interestingly, the S4b architecture significantly enhanced cell migration speed, persistence, and MSD. The average cell migration speed is $0.231 \pm 0.013$ μm/min for S4b in contrast to $0.160 \pm 0.007$ μm/min for S2b, which is a 44.2% difference. The average persistence is $0.638 \pm 0.024$ for S4b in contrast to $0.403 \pm 0.021$ for S2b, or a 58.3% difference. When we compared the speed, persistence and MSD of all four cases, it was clear that spheroids with architecture S4b had highest speed, persistence and MSD.

Taken together, we found that MDA-MB-231 cells from co-culture spheroid of S4b architecture to be the most invasive. This invasiveness was largely due to the architectural arrangement of the two cells types, where the malignant MDA-MB-231 cells were surrounded by the non-tumorigenic MCF-10A cells with an opening at the peripheral. This opening enabled the MDA-MB-231 cells to migrate much more directionally than other spheroid architectures. It is interesting to note that although malignant tumor cells in day 4 spheroids were not readily detached from the spheroids initially (in comparison to cells from spheroids of day2), they were much more invasive once they invaded into the ECM.

**Conclusion and future perspectives**

Here, we showed that the architectural arrangement of the malignant tumor cells and the non-tumorigenic epithelial cells critically regulated tumor cell invasiveness. One important finding was that the non-tumorigenic epithelial shell can prevent the invasion of the malignant tumor cells when they were surrounded by the non-tumorigenic cells. Among the four tumor architectures we created, we found that S4b, the case where tumor cells were surrounded by the non-tumorigenic cells with a small opening to the ECM. This architecture facilitated a persistent and fast movement of the malignant cells. In all four spheroid architectural cases, the invading cells were those that had direct contact with the ECM.

We propose a 3D co-culture tumor spheroid invasion model here for the studies of cell-cell interactions in tumor invasion. Our work highlighted the importance of tissue architecture in tumor invasion. The tumor spheroid assay presented here is a step forward in comparison to the single cell invasion assays [27]. It provided a straightforward platform to include cell-cell interactions within the tumor bulk in tumor invasion. Future direction will be to include immune cells, and other stroma cells within the tumor environment. This work can also be easily adapted to other cancer types.

Looking forward, an important question is how the tumor microenvironment regulates tumor architecture and subsequent invasion. Clinically, the transition from ductal carcinoma in situ (DCIS) to invasive ductal carcinoma (IDC) for breast cancer patients is still poorly understood (Fig. S1A) [4]. We note the architectural similarities between DCIS to S4a spheroids (Fig. S1B), and IDC to S4b spheroids (Fig. S1C). We believe that the presented tumor spheroid assay can potentially be used for modeling different stages of breast cancer, which will facilitate an understanding of the transition from DCIS to IDC. An immediate question will be the roles of biophysical and biochemical cues within the tumor environment in facilitating the transition from DCIS to IDC and subsequently tumor invasion.

## Materials and Methods

Cell culture: Metastatic breast cancer cells (MDA-MB-231 cell line expressing EGFP) and non-tumorigenic mammary epithelial cells (MCF-10A cell line expressing dTomato variants) were kind gifts from Dr. Joseph Aslan at the Oregon Health & Science University. MDA-MB-231 cells were cultured every 3 to 4 days from passage 2 to 20, and used at 50-70% confluency [28]. The growth medium for MDA-MB-231 cells was composed of DMEM high glucose medium (Gibco, Life Technologies Corporation, Grand Island, NY), 10% fetal bovine serum (Atlanta biologicals, Lawenceville, GA), and 1% antibiotics (100 units/mL penicillin and 100 µg/mL streptomycin). MCF-10A cells were cultured every 3 to 4 days from passage 2 to 10, and used at 70-90% confluency. The growth medium for MCF-10A cells was composed of DMEM/F-12 medium (Gibco), 5% donor horse serum (Atlanta biologicals), 20 ng/mL human EGF (Gibco), 0.5 µg/mL hydrocortisone (Sigma-Aldrich, St. Louis, MO), 100 ng/mL Cholera Toxin (Sigma-Aldrich), 10 µg/mL insulin (Sigma-Aldrich), and 1% antibiotics (Gibco).

Co-culture spheroid formation: Tumor spheroids were formed using microfabricated microwell arrays previously developed in our labs [21, 22]. Each spheroid was formed within a 200 µm diameter and 220 µm height non-adherent microwell treated with 1% pluronic F-127 solution (Sigma-Aldrich). The microwell array was patterned on a thin PDMS membrane (Fig. S2A) and then glued to the bottom surface of one of the 12-well plates. Each microwell array contained 1296 microwells (Fig. S2A), and within each microwell array, 2 million cells (1:1 ratio of EGFP MDA-MB-231: dTomato MCF-10A) suspended in 2.5mL medium (1:1 ratio of DMEM and DMEM/F12 growth media) were seeded. Cells were first allowed to settle down into all microwells for 30 minutes in the incubator (Fig. 1A) before the device was placed on a rocker (Boekel Scientific, Rocker II Model 260350). Co-cultured spheroids were formed within the microwells after overnight and cultured for 2 or 4 days to obtain different architectures (Fig. S2B), with medium change every 2-3 days. The diameter of the spheroids was approximately 100 µm to fit the invasion device height constraint of 200 µm.

Experimental procedure: A microfluidic platform previously developed in our lab was used to study the co-culture spheroid invasion in 3D collagen matrices [25, 26]. Briefly, sterilized PDMS devices were treated with oxygen plasma (Harrick Plasma Cleaner PDC-001, Harrick Plasma, Ithaca, NY) for 1 minute on high power mode and then activated with 1% Poly(ethyleneimine) (Sigma-Aldrich) for 10 minutes followed by a 0.1% Glutaraldehyde (Electron Microscopy Sciences, Hatfield, PA) treatment for 30 minutes. The PDMS device was then sandwiched between a 3 inch by 1 inch glass slide and a plastic manifold. A 0.6% of agarose solution was used to fill the void space around the PDMS device to prevent medium from evaporating during spheroid invasion experiment.

On the day of experiment, the spheroids were collected from the microwells and filtered by a Falcon® Cell Strainer (Corning) with 70µm pores to ensure the uniformity of the spheroid size for each experiment. To prepare spheroid embedded collagen matrices, 60 µL Type I collagen from stock concentration of 5.0 mg/mL (Corning, Discovery Labware Inc., Bedford, MA) was first titrated with 1.32µL 1N NaOH and 20 µL 10X M199 (Sigma-Aldrich) to yield a final pH of approximately 7.4 [Cross et al. 2010]. The collagen was then mixed with the collected co-culture spheroids to a final volume of 200 µL. The final average spheroid concentration was approximately one spheroid per one $mm^2$ from the top view of the device and the final collagen concentration was 1.5 mg/mL.

Spheroid-embedded collagen solution was introduced to all the channels in each device on an ice block. The microfluidic device was then placed in an incubator to allow collagen polymerization for 45 minutes at 37°C. To prevent spheroids from gravitationally settling down to the bottom of the device, the microfluidic chips were positioned up-side-down for the first 10 minutes and then flipped three times more at time points 5, 15, and 15 minutes in the incubator. Following polymerization, the channels and reservoirs were hydrated with 37 °C 1:1 ratio medium, then were plugged with PDMS filled gel loading tips. The microfluidic device was then transferred to the microscope stage enclosed by an environmental control chamber (WeatherStation, PrecisionControl LLC), which was kept at 37 °C, 5% CO2 and about 70% humidity).

Imaging and data analysis: An inverted microscope (IX81, Olympus America, Center Valley, PA, USA) with a CCD camera (Orca-ER, Hamamatsu Photonics, Japan) was used for all the invasion experiments. In a typical experiment, the middle z-plane of the spheroids in the channels were captured using a 10X objective (Olympus, NA=0.3) in bright field mode and in green fluorescence mode (EX:460-500nm, EM: 510-560nm) for EGFP-MDA-MB-231 cells and red fluorescence (EX: 510-560nm, EM: 572.5-647.5nm) mode for dTomato-MCF-10A cells. A sequence of 109 images was captured every 20 minutes for a total of 36 hours.

To quantify tumor spheroid invasion, fluorescent images of the GFP mode were used to quantify tumor cell invasion by calculating the percentage of cells detached from the spheroid. Here, fluorescence intensity of the image was used to represent the spatial cell density. To determine the percentage of MDA-MB-231 cells detached from the spheroid, an outline of the spheroid was drawn around the tumor edge in bright field image at t = 0 using ImageJ. Then the outline was superimposed to the GFP fluorescent image of the same spheroid (Fig. S3). The total intensity within the outline was measured using ImageJ at all time points in 36 hours range. The percentage of cells detached from the spheroid was calculated by the ratio of the total fluorescence outside the initial spheroid outline divided by the total fluorescence of the spheroid at t=0.

To quantify tumor cell motility, time-lapse images of tumor spheroid invasion were processed in ImageJ and in house Matlab programs. MDA-MB-231 tumor cells were tracked after they invaded out of the spheroid. Note that the starting time for each cell differed because each cell invaded out from the spheroid at different time points. A total of 60 tumor cells were tracked for each spheroid architecture and all the tracked trajectories were used to compute the cell migration speed, persistence, and the mean square displacements (MSDs) [22, 28]

Statistical Analysis. All the data were plotted using Matlab or Prism GraphPad software. Student's t-test and one-way ANOVA test were performed for two-group and four-group comparisons, respectively using Prism and mean ± SEM were presented in all numerical results as well as the average line and error bars in the plots.

## Acknowledgments

This work was mainly supported by a grant from the National Cancer Institute [Grant no. R01CA221346]; partially supported by the Cornell Center on the Microenvironment & Metastasis [Award No U54CA143876 from the National Cancer Institute]; the Cornell NanoScale Science and Technology, and the Cornell BRC imaging facility.  We thank Professor Mingling Ma's lab for sharing their protocols for making spheroids and Young Joon Suh for helpful discussions.

# Figures and captions

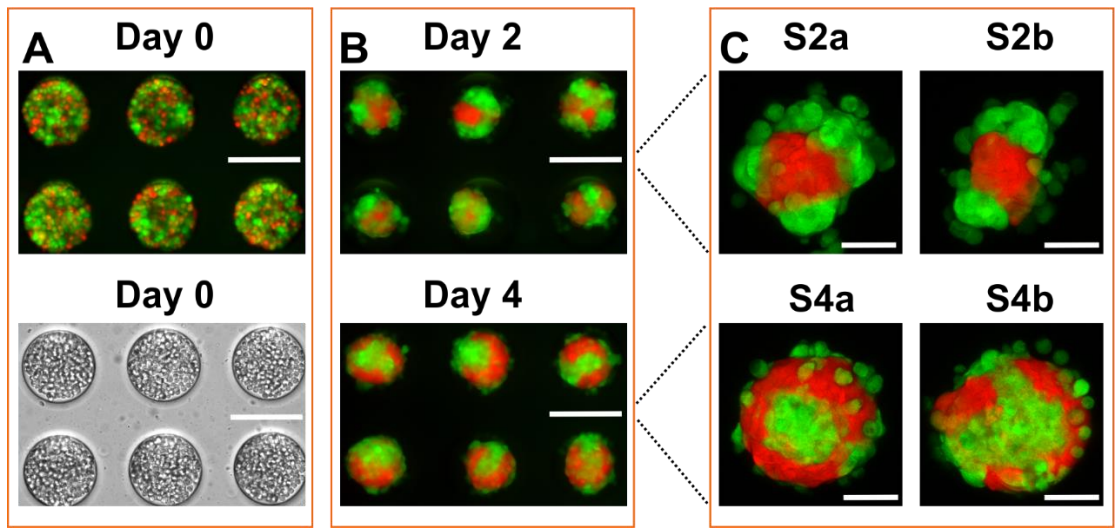

**Figure 1: Different architectures of co-culture spheroids formed within an array microwell platform.** Co-culture tumor spheroids were formed by mixing two cell types in an array microwell. **A.** Fluorescence (top panel) and bright field (bottom panel) micrographs of 1:1 ratio of two cell types in microwells at day 0. **B.** Florescence image of co-culture spheroids at day 2 and day 4. **C.** Close-up confocal images of four types of spheroid architecture observed at day 2 and day 4. Spheroids on day 2 exhibited a morphology with metastatic MDA-MB-231 cells outside and nontumorigenic MCF-10A cells inside (S2a and S2b). Spheroids on day 4 presented a reversed morphology with MCF-10A cells surrounding MDA-MB-231 cells, with a few loosely MDA-MB-231 cells attaching the periphery (S4a and S4b). In S2a, and 4a, one cell type were enclosed by the other cell type. In S2b, 4b, one cell type only partially surrounded by the other cell type. Green cells were malignant breast tumor cell line (MDA MB-231) expressing green fluorescent protein; red cells were a non-tumorigenic breast epithelial cell line (MCF-10A) expressing dTomato. Scale bar: 200 µm in A and B and 50 µm in C.

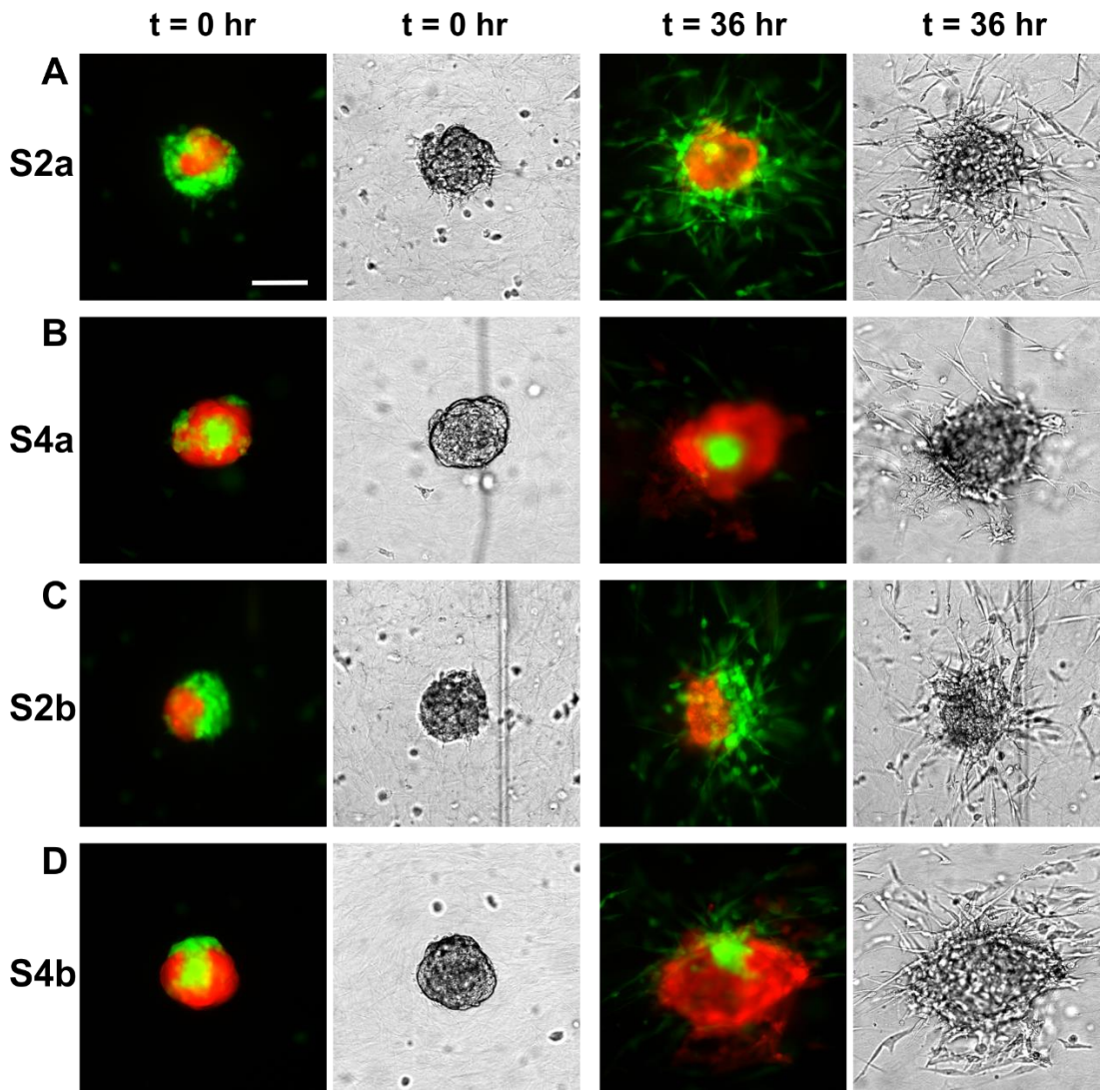

**Figure 2: Architecture of the co-culture spheroids regulated tumor invasion in 3D collagen matrices.** Fluorescence (left) and brightfield (right) images of tumor spheroids at time points t = 0 and 36 hrs. **A.** In the spheroids with MDA-MB-231 cells outside of a MCF-10A cluster (S2a), significant numbers of the MDA-MB-231 cells invaded out of the spheroid. **B.** In the spheroids with MCF-10A cells outside of MDA-MB-231 cell cluster (S4a), only the peripheral MDA-MB-231 cells invaded out of the spheroid while the majority malignant MDA-MB-231 cells enclosed within the MCF-10A cells remained inside. **C.** In the spheroids with two cell types side by side (S2b), significant numbers of MDA-MB-231 cells invaded out of the spheroid from one side of the spheroid. **D.** In the spheroids with MCF-10A surrounding MDA-MB-231 cells with an opening (S4b), MDA-MB-231 cells invaded out through the opening. The collagen concentration is 1.5 mg/mL. Scale bar: 100 μm.

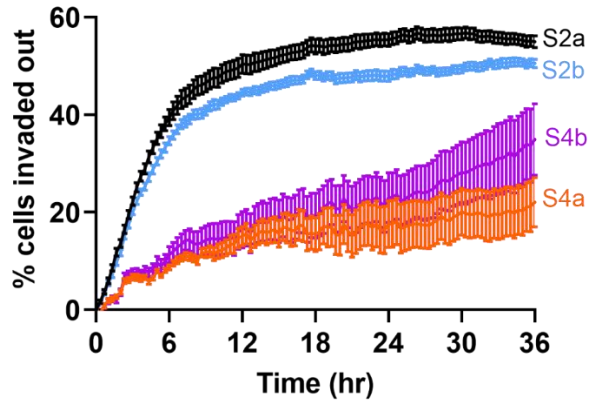

**Figure 3: Tumor cell detachment from spheroids** Percentage of tumor cells (MDA-MB-231 cells) detached from the co-culture spheroids of four different architecture. Time lapse images were used to calculate the fluorescence of the MDA-MB-231 cells. Percentages of detaching cells for spheroids S2a, S2b, S4a, and S4b are 55.0 ± 1.19 %, 50.5 ± 0.88 %, 22.1 ± 5.07 %, and 34.9 ± 7.34 %.

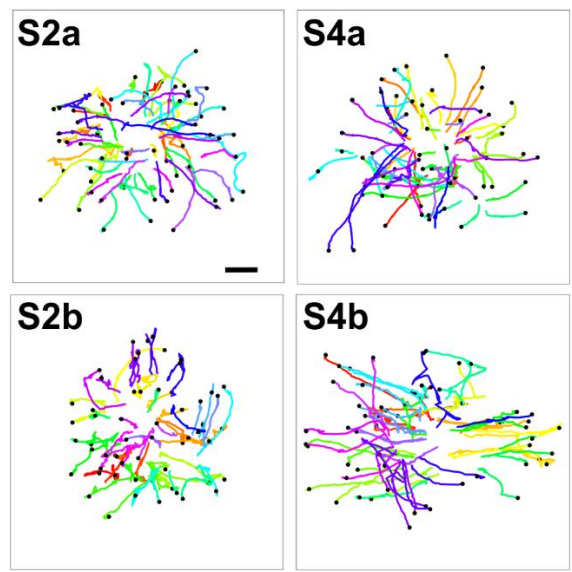

**Figure 4: Trajectories of MDA-MB-231 cells invading out of the tumor spheroids.** 60 tumor cell trajectories were presented for each spheroid architecture. Each colored line represents one cell trajectory. The start point of each trajectory was at the x and y coordinate for the cell position when it started to migrate away from the spheroid, and the end point is presented as a black dot. The time durations of the trajectories range from 14 to 36 hours (average of 28 hours) for S2a, 10 to 36 hours (average of 29 hours) for S4a, 12 to 36 hours (average of 29 hours) for S2b and 6 to 36 hours (average of 23 hours) for S4b.

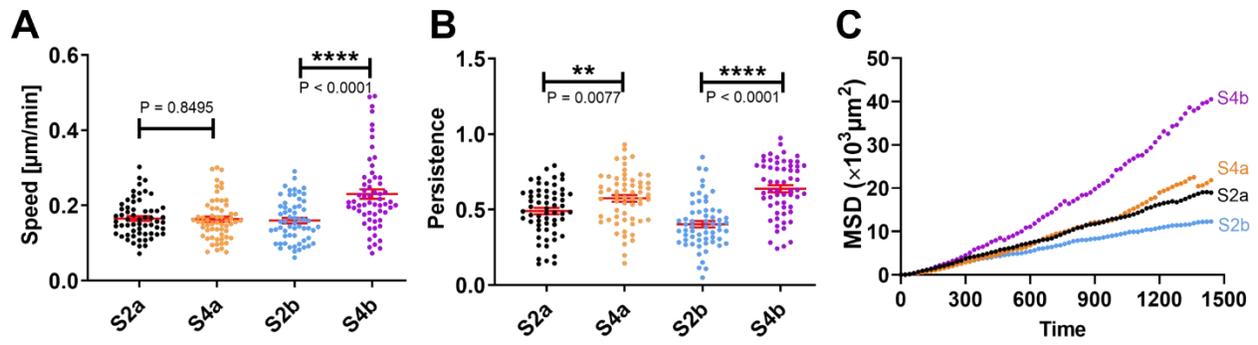

**Figure 5: Architecture of spheroids modulated speed, persistence, and mean squared displacements (MSDs) of MDA-MB-231 malignant breast tumor cells. A.** Migration speed (in μm/min) of invading MDA-MB-231 tumor cells. S4b has a significant increase in cell migration speed in comparison to that of S2b. **B.** Persistence of invading MDA-MB-231 tumor cells. Persistence of invading cells from day 4 spheroids is higher than those from day 2. One-way ANOVA test was used for both speed and persistence comparison for all four architecture with p<0.0001. **C.** Mean squared displacements (MSDs) of invading MDA-MB-231 tumor cells. Tumor cells from S4b have significantly greater MSD than those of the other architectures in the time duration of 24 hours. For each architecture, n = 60 (4 co-culture spheroids, with 15 cells each were analyzed). At least 30 cells for each condition were used for the MSD calculation.

**Supplementary materials:**

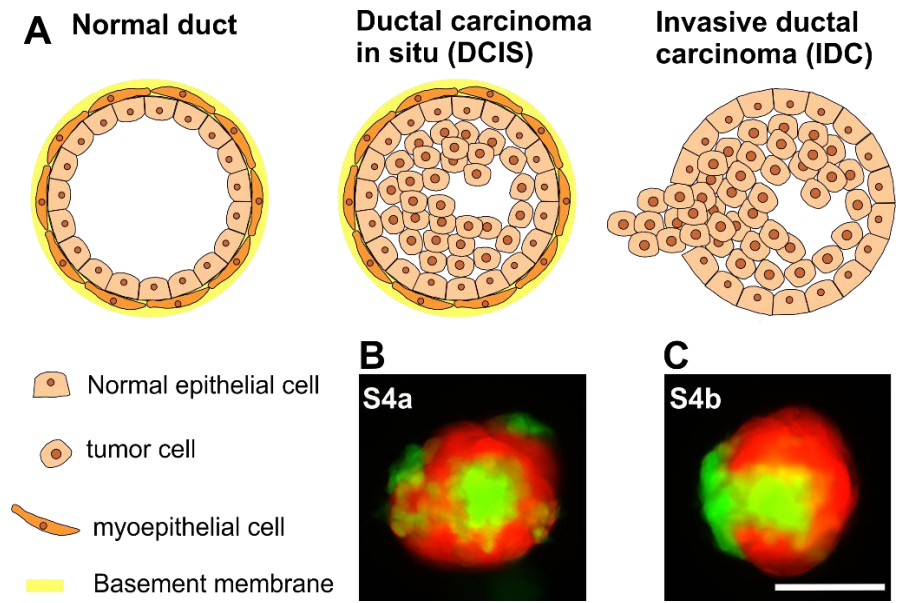

**Figure S1: Modeling different stages of breast cancer.** A. Progression of breast cancer from a normal duct to ductal carcinoma in situ (DCIS) to invasive ductal carcinoma (IDC). B. A co-culture spheroid with an architecture of non-tumorigenic epithelial cells (red) enclosing malignant tumor cells (green), mimicking DCIS. C. A co-culture spheroid with an architecture of non-tumorigenic epithelial cells (red) partially surrounding the malignant tumor cells (green) with an opening, mimicking IDC.

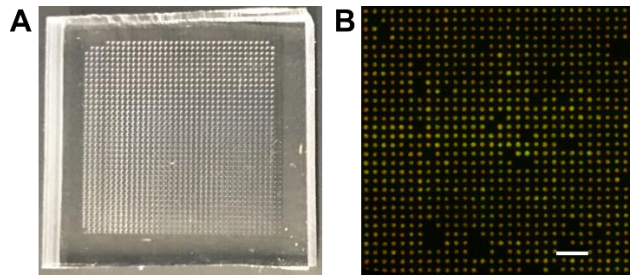

**Figure S2: Microwell array platform for spheroid formation.** A. PDMS membrane with an array of 36 × 36 microwells. Each well has a diameter of 200 µm and a height of 220 µm. B. Fluorescent images of the co-culture spheroids formed within the microwells. Each color dot is one spheroid. Scale bar is 1 mm.

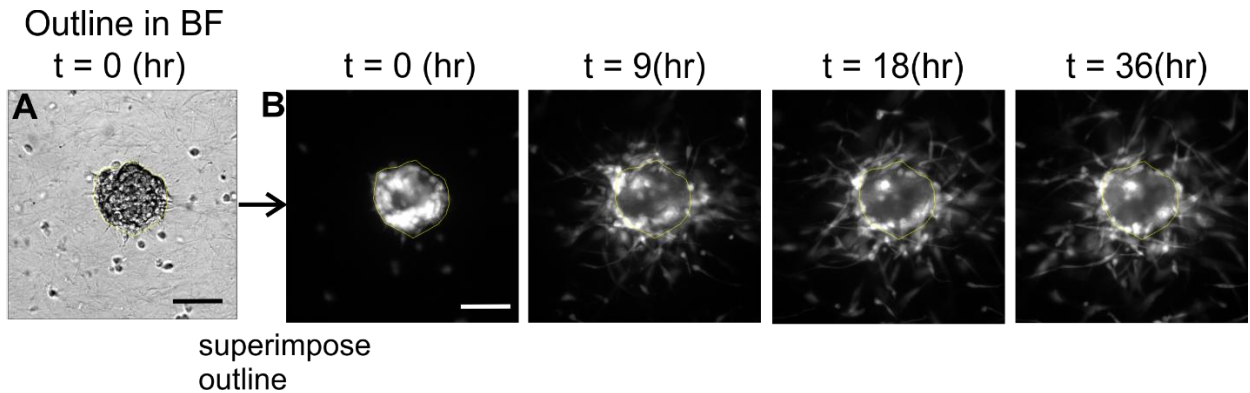

**Figure S3: Quantification of the percentage of tumor cells invading out of the co-culture spheroid.** A. Brightfield image of a co-culture spheroid with a yellow outline at the periphery at t = 0 hour. B. The outline was then superimposed to the GFP fluorescent images of the MDA-MB-231 cells at t = 0, 9, 18, and 36 hours. Scale bar is 100 µm for both A and B.

**Movie Legends:**

Movie 1: **S2a architecture co-culture tumor spheroid invasion within 1.5mg/mL collagen.** Each image is 451.5 µm × 451.5 µm, the time between consecutive image is 20 minutes, and the duration of the move is 36 hours. Green: MDA-MB-231 cells. Red: MCF-10A cells.

Movie 2: **S4a architecture co-culture tumor spheroid invasion within 1.5mg/mL collagen.** Each image is 451.5 µm × 451.5 µm, the time between consecutive image is 20 minutes, and the duration of the move is 36 hours. Green: MDA-MB-231 cells. Red: MCF-10A cells.

Movie 3: **S2b architecture co-culture tumor spheroid invasion within 1.5mg/mL collagen.** Each image is 451.5 µm × 451.5 µm, the time between consecutive image is 20 minutes, and the duration of the move is 36 hours. Green: MDA-MB-231 cells. Red: MCF-10A cells.

Movie 4: **S4b architecture co-culture tumor spheroid invasion within 1.5mg/mL collagen.** Each image is 451.5 µm × 451.5 µm, the time between consecutive image is 20 minutes, and the duration of the move is 36 hours. Green: MDA-MB-231 cells. Red: MCF-10A cells.